\documentclass{Interspeech}
\usepackage{adjustbox}

\interspeechcameraready

\title{A Two-Stage Hierarchical Deep Filtering Framework for Real-Time Speech Enhancement}

\author[affiliation={1}]{Shenghui}{Lu}
\author[affiliation={1}]{Hukai}{Huang}
\author[affiliation={1}]{Jinanglong}{Yao}
\author[affiliation={1}]{Kaidi}{Wang}
\author[affiliation={1}]{Qingyang}{Hong$^*$}
\author[affiliation={2}]{Lin}{Li$^*$}

\affiliation{School of Informatics}{Xiamen University}{China}
\affiliation{School of Electronic Science and Engineering}{Xiamen University}{China}
\email{lushenghui@stu.xmu.edu.cn}

\usepackage{comment}

\begin{document}
\maketitle

\renewcommand{\thefootnote}{\fnsymbol{footnote}}
\footnotetext[1]{\hspace{-0.5em}$^*$Corresponding author}
\renewcommand{\thefootnote}{\arabic{footnote}}

\begin{abstract}
This paper proposes a model that integrates sub-band processing and deep filtering to fully exploit information from the target time-frequency (TF) bin and its surrounding TF bins for single-channel speech enhancement. The sub-band module captures surrounding frequency bin information at the input, while the deep filtering module applies filtering at the output to both the target TF bin and its surrounding TF bins. To further improve the model performance, we decouple deep filtering into temporal and frequency components and introduce a two-stage framework, reducing the complexity of filter coefficient prediction at each stage. Additionally, we propose the TAConv module to strengthen convolutional feature extraction. Experimental results demonstrate that the proposed hierarchical deep filtering network (HDF-Net) effectively utilizes surrounding TF bin information and outperforms other advanced systems while using fewer resources.
\end{abstract}

\section{Introduction}

In recent years, deep learning methods have achieved remarkable success in single-channel speech enhancement \cite{deepfilternet,cmgan,fullsubnet,gtcrn}. Mainstream speech enhancement approaches typically apply the short-time Fourier transform (STFT) to convert noisy speech signals from the time domain to the frequency domain, where a deep neural network estimates a spectral mask for each TF bin to obtain the clean speech spectrum. The enhanced speech signal is then reconstructed by applying the inverse STFT (iSTFT).

Recent studies have revealed that the TF bins surrounding a target TF bin in the spectrogram play a crucial role in accurately predicting its mask \cite{fullsubnet,deepfiltering}. One approach that leverages surrounding TF bins is FullSubNet \cite{fullsubnet}, which processes each frequency independently in its sub-band model. The input consists of the target TF bin and its surrounding frequency bins. FullSubNet enhances input features using this method, achieving impressive results. However, sub-band methods focus primarily on local spectral features and cannot capture long-range spectral dependencies, requiring additional extraction of full-band features. Another approach that models surrounding TF bins is deep filtering \cite{deepfiltering}, which replaces the one-to-one spectral mask with a multi-to-one filter that leverages information from surrounding TF bins. However, deep filtering faces challenges in predicting filter coefficients accurately due to the absence of ground truth \cite{deepfiltering}, especially when the filter order is large, which may result in some performance degradation \cite{dforderbigworse}.

\begin{figure}[t]
\centering
\includegraphics[width=0.47\textwidth, trim=6.25cm 1.13cm 1.13cm 2.2cm, clip]{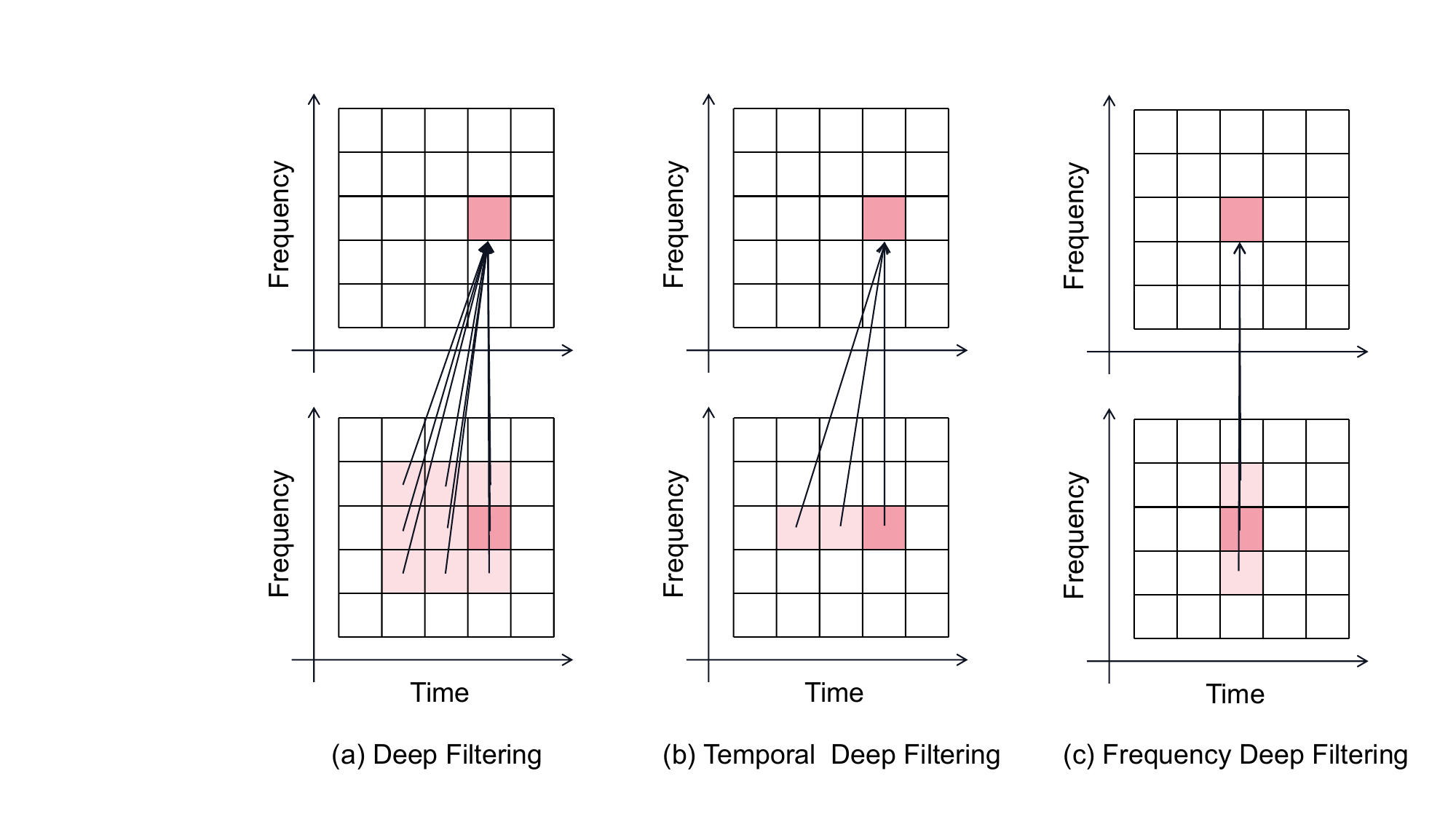} 
\caption{Illustration of deep filtering, temporal deep filtering, and frequency deep filtering with order=3.}
\label{fig:fig1}
\end{figure}

Recently, multi-stage speech enhancement networks have thrived by decomposing the single-step prediction problem into simpler subproblems \cite{atwostage,twoheadsbetterthanone,firstcoarsefineafterward,snr1,snr2}, which generally lead to better performance. DeepFilterNet \cite{deepfilternet} employs deep filtering in the time domain to enhance periodic components and achieves better results. Therefore, decoupling the time and frequency domains provides an effective strategy for mitigating the complexity of predicting deep filtering coefficients. Moreover, some studies have shown that deep filtering in the time domain alone outperforms filtering in the frequency domain \cite{deepfiltering,dforderbigworse}, suggesting that capturing more information from the time domain, rather than the frequency domain, is more effective in improving speech enhancement model performance.

\begin{figure*}[t]
\centering
\includegraphics[width=1\textwidth, trim=0cm 3.0cm 0cm 0.9cm, clip]
{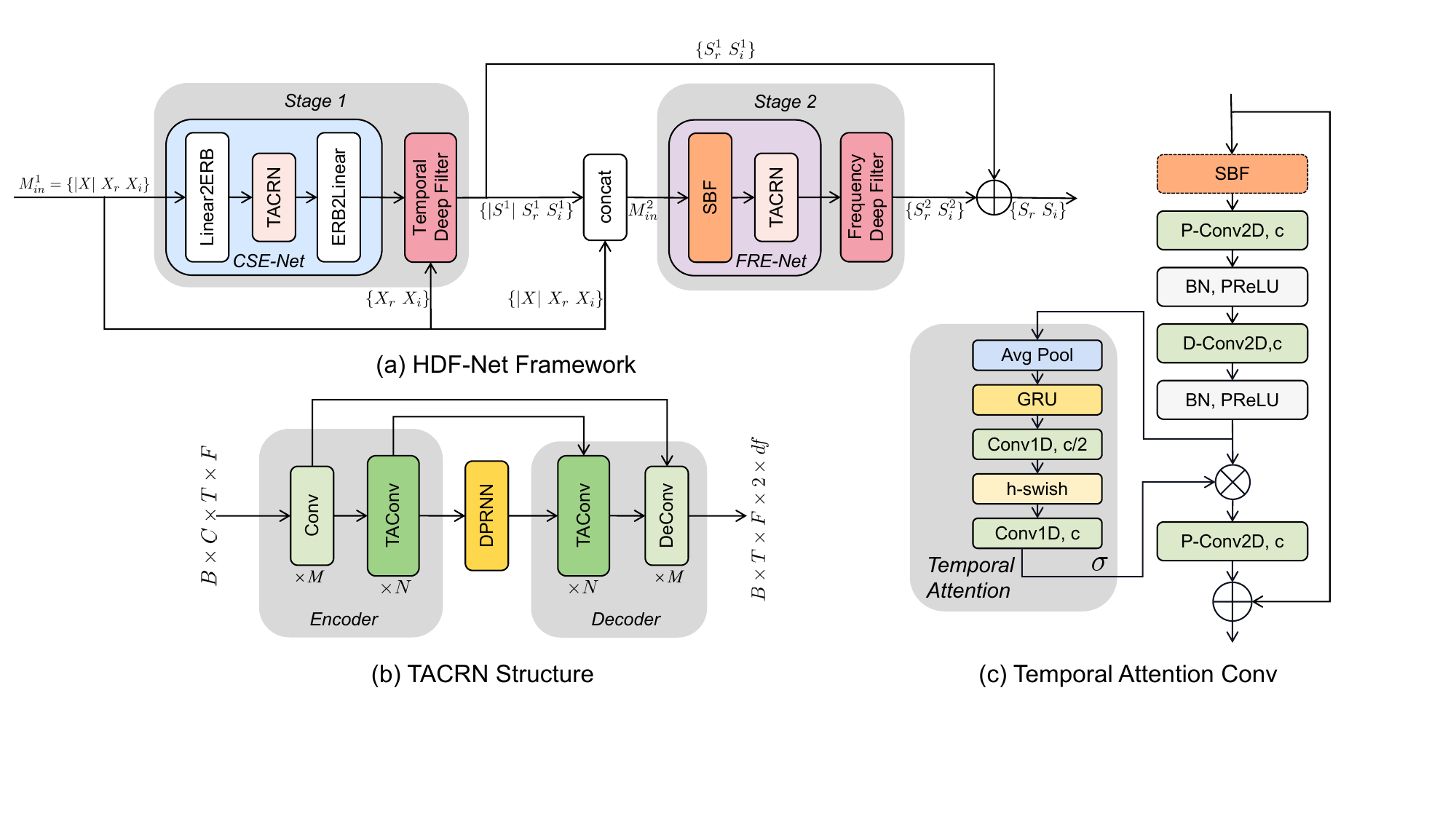} 
\caption{ (a) Framework of the Proposed HDF-Net; (b) Detailed structure of TACRN; (c) Details of the proposed temporal attention conv (TAConv) module. The dashed box contains the optional SBF module, which is used only in the FRE-Net during the second stage.}
\label{fig:framework}
\end{figure*}
Motivated by the issues discussed above, we propose a two-stage speech enhancement network, the hierarchical deep filter network (HDF-Net), consisting of two main components: the coarse spectral enhancement network (CSE-Net) in the first stage and the fine residual enhancement network (FRE-Net) in the second stage. In the first stage, a band-based partitioning approach is used to expand the convolutional receptive field in the frequency domain for coarse spectral processing, followed by temporal deep filtering (as shown in Figure \ref{1}(b)) to enhance periodic components of speech. In the second stage, a sub-band method is applied to model local spectral features more accurately and predict the filtering coefficients
, enabling frequency deep filtering (as shown in Figure \ref{1}(c)). Additionally, we introduce a lightweight convolution module, temporal attention conv (TAConv), which utilizes pointwise convolution \cite{pconv} and depthwise convolution to reduce computational complexity, while a temporal attention module captures energy-related information along the temporal axis. 
Overall, our contributions in this paper can be summarized as follows.
\begin{itemize}
    \item We integrate the sub-band processing and deep filtering, leveraging information from surrounding TF bins to improve the model performance.
    \item We decouple the temporal and frequency domains of deep filtering and adopt a two-stage strategy to effectively reduce the burden of predicting filter coefficients at each stage.
    \item To improve the feature extraction capability of convolutional modules, we propose a lightweight convolutional module called TAConv, designed to capture temporal correlations.
\end{itemize}

We evaluated the proposed HDF-Net on the VoiceBank+DEMAND datasets, achieving superior performance with a relatively low parameter count.

\section{Methodology}
\subsection{The overall architecture}

The noisy speech signal is denoted as $x(n)$, which can be expressed as:
\begin{equation}
x(n)=s(n)+z(n) \label{1}
\end{equation}
where $s(n)$ is clean speech signal, $z(n)$ is additive noise signal. Typically, noise reduction operates in frequency domain:
\begin{equation}
X(t,f) = S(t,f) + Z(t,f) \label{2}
\end{equation}
where $X(t, f)$ represents the signal $x(n)$ in the frequency domain, $t$ and $f$ denote the time and frequency bins respectively.

Our HDF-Net consists of two stages, as illustrated in Figure \ref{fig:framework}(a). The input $M_{in}^1$ of stage 1 contains the magnitude spectrum $|X|$, the complex spectrum $X_r$ and  $X_i$, $M^1_{in} \in \mathbb{R}^{B \times 3\times T\times F}$.
In the first stage, we use CSE-Net to predict the temporal filtering coefficients and apply temporal deep filtering (Section \ref{section:tdf}) for initial speech enhancement, resulting in the complex spectrograms $S_r^1$ and $S_i^1$. 

To provide more comprehensive input, the second-stage input $M^2_{in}$ is not only the output from the first stage but also concatenates the original noisy speech spectrum:
\begin{equation}
M^2_{in}=Concat(|X|,X_i,X_r,|S^1|,S^1_i,S^1_r) 
\label{2}
\end{equation}
where $M^2_{in}\in \mathbb{R}^{B \times 6\times T\times F}$.
In the second stage, we use FRE-Net to predict frequency filtering coefficients, followed by frequency deep filtering (Section \ref{section:fdf}) to perform fine-grained local enhancement of the spectrogram, yielding $S_r^2$ and $S_i^2$.

Rather than directly outputting the predicted spectrogram in stage 2, we employ a residual approach \cite{twoheadsbetterthanone}, where the second-stage model output corrects the first-stage output. The final predicted spectrogram $S$ is the sum of the outputs from both stages:
\begin{equation}
\begin{cases}
S_r = S^1_r + S^2_r \\
S_i = S^1_i + S^2_i
\end{cases}
\label{2}
\end{equation}

\subsection{Temporal and Frequency Deep Filtering}
\subsubsection{Deep Filtering}

Deep filtering was first proposed by Mack et al. \cite{deepfiltering}. It is a filter that operates in the TF domain:
\begin{equation}
S(t,f) =\sum\limits_{i=0}^{I} \sum\limits_{j=-J}^{J} C_{tf}(t,i,f,j)\cdot X(t-i,f-j) \label{equation:df}
\end{equation}
where $C_{tf}$ is the complex coefficients of the deep filter applied to the input spectrum $X$. $S$ is the enhanced spectrum. The magnitude of $I$ and $J$ control the order of the deep filter in the TF domains. The value of $i$ is constrained to be non-negative to ensure causality. In this paper, the deep filtering is decoupled into temporal deep filtering and frequency deep filtering, as shown in Figure \ref{1}.

\subsubsection{Temporal Deep Filtering}
\label{section:tdf}
Temporal deep filtering is a filtering method that operates only in the time domain:
\begin{equation}
S(t,f)=\sum\limits_{i=0}^{I} C_t(t,i,f)\cdot X(t-i,f) \label{2}
\end{equation}
where $C_t$ is the complex coefficient of the temporal deep filter applied to the input spectrum $X$. The temporal deep filtering for the current frame is obtained by filtering the spectra of the current frame and the preceding $I$ frames, with an effect similar to the comb filter used in RNNoise \cite{rnnoise} to reduce noise within harmonic components.
\subsubsection{Frequency Deep Filtering}
\label{section:fdf}
Frequency deep filtering is a filtering method that operates only in the frequency domain:
\begin{equation}
S(t,f)=\sum\limits_{j=-J}^{J} C_f(t,f,j)\cdot X(t,f-j) \label{2}
\end{equation}
where $C_f$ is the complex coefficients of the frequency deep filter applied to the input spectrum $X$. Frequency deep filtering obtains the result for the current TF bin by filtering the surrounding frequency bins, focusing on modeling local spectral features.

\subsection{CSE-Net and FRE-Net}
As shown in CSE-Net in Figure \ref{fig:framework}(a), to reduce computational complexity and expand the convolutional receptive field, we utilize ERB \cite{erb} frequency bands to transform linear frequencies into ERB frequencies.
As shown in  FRE-Net in Figure \ref{fig:framework}(a). We use a sub-band fusion (SBF) module to extract local frequency information \cite{fullsubnet}:
\begin{equation}
X_{sbf} = SBF(X) \label{2}
\end{equation}
where, $X\in \mathbb{R}^{B \times 3\times T\times F}$, $X_{sbf}\in \mathbb{R}^{B \times (3\times k)\times T\times F}$. SBF module replicates the $k$ bands around each band into the channel dimension, enhancing the sub-band information. In the second stage, the subsequent frequency deep filtering further exploits the extracted sub-band information.


CSE-Net and FRE-Net use the convolutional recurrent network (CRN) \cite{crn} structure, as shown in Figure \ref{fig:framework}(b). The CRN we use, called TACRN, consists of a symmetric encoder-decoder structure with convolution (Conv) and temporal attention convolution (TAConv) blocks.
The Conv blocks are responsible for downsampling the frequency dimension, while TAConv blocks further extract features (Section \ref{section:TAConv}). The deconvolution (DeConv) block contains the same components as the Conv block, except that the convolution layer is replaced with a transposed convolution layer. The network outputs the deep filter complex coefficients, with a shape of $B\times T\times F\times 2\times df$, where $df$ represents the filter order. Skip connections are applied between the corresponding blocks of the encoder and decoder to reduce information loss during the encoding process.

As shown in Figure \ref{fig:framework}(b), we use DPRNN \cite{dprnn} as the RNN module in the TACRN. DPRNN has been shown to be a very effective modeling approach for speech enhancement \cite{gtcrn,dpcrn}.
It sequentially models both intra-frame and inter-frame dependencies,
Specifically, the input feature map of shape $B \times C \times T \times F$ is first reshaped as $BT \times F \times C$. 
The intra-frame frequency dependence is extracted via a bi-directional GRU, which is then reshaped as $BF \times T \times C$. 
The temporal dependence on the frequency is extracted by a one-way GRU.  In order to reduce the complexity of the model to ensure real-time performance, we use the grouped GRU \cite{gtcrn}.

\subsection{Temporal Attention Conv}
\label{section:TAConv}
We design the TAConv for speech enhancement, as shown in Figure \ref{fig:framework}(c). First, an optional SBF module extracts sub-band features, which is applied only in the FRE-Net during the second stage. Next, pointwise conv2d (P-Conv2D) \cite{pconv} and depthwise conv2d (D-Conv2D) with a 3×3 kernel size are employed to extract TF features with reduced complexity, replacing conventional convolutions. Inspired by TF attention \cite{tfattention}, we propose a causal temporal attention (TA) module. 
Specifically, for the input feature tensor with shape $B \times C \times T \times F$, we first apply average pooling along the frequency dimension to obtain a tensor of shape $B \times C \times T$. After extracting temporal information with a GRU, we apply squeeze-and-excitation \cite{sequeezeandexcitation}. Since surrounding time frames also provide valuable information for speech enhancement, we replace the fully-connected layer with a Conv1D module. Subsequently, we apply a sigmoid function to compute channel weights and apply them to the extracted features. Finally, a P-Conv2D layer is applied, and the residual connection is used to obtain the final output.

\subsection{Loss Function}
We optimize the model by minimizing both the magnitude spectrum and the complex spectrum loss.:
\begin{equation}
\mathcal{L} = \alpha \mathcal{L}_{\text{mag}}(S, \tilde{S}) + \beta \mathcal{L}_{\text{comp}}(S, \tilde{S}) \label{2}
\end{equation}
where $S$ and $\tilde{S}$ are the spectra of the target and predicted speech, respectively, while $\alpha$ and $\beta$ are the loss weights. $\mathcal{L}_{\text{mag}}$ and $  \mathcal{L}_{\text{comp}}$ denote the amplitude spectrum and complex spectral loss:
\begin{equation}
\mathcal{L}_{\text{mag}}(S,\tilde{S}) = \text{MSE}(|S|^{c},|\tilde{S}|^{c})
\label{2}
\end{equation}
\begin{equation}
\mathcal{L}_{\text{comp}}(S,\tilde{S}) = \text{MSE}(S_r^{c},\tilde{S}_r^{c}) + \text{MSE}(S_i^{c},\tilde{S}_i^{c})
\label{2}
\end{equation}
where $S_r$ and $S_i$ denote the real and imaginary parts of the spectrum of the target speech $S$, respectively. To make the enhanced speech more in line with human hearing, we use $c$ as the power to compress the amplitude and complex spectrum, which helps the model better concentrate on the low-energy components. In this paper, we set $c=0.3$, and the loss of the complex spectrum $\mathcal{L}_{comp}$ includes the loss for both the real and imaginary parts. The mean squared error (MSE) is used to compute this loss.

Similar to \cite{twoheadsbetterthanone}, we also use a two-stage training approach to optimize our HDF-Net. Specifically, initially train the first-stage network until convergence. Then, the first and second stages are jointly trained, with the above losses used in both training phases.

\section{EXPERIMENTS AND RESULTS}
\subsection{Datasets}
To evaluate the performance of HDF-Net, we conducted experiments on the VoiceBank+DEMAND \cite{voicebank_demand} dataset, which contains paired clean and noisy speech. The training set consists of 11,572 utterances from 28 speakers, with signal-to-noise ratios (SNRs) of {0, 5, 10, 15} dB. The test set includes 824 utterances from 2 unseen speakers, with SNRs of {2.5, 7.5, 12.5, 17.5} dB. All utterances were re-sampled to 16 kHz, and during training, speech segments were randomly cropped to 2 seconds.
\subsection{Implementation Details}
In our experiment, we applied STFT with a Hanning window, using a window length of 32 ms, a hop size of 16 ms (50\% overlap), and a 512-point FFT. Similar to \cite{gtcrn}, we configured the ERB frequency bands by keeping 65 low-frequency bands fixed while mapping 192 high-frequency bands to 64 ERB bands, resulting in a total of 129 compressed frequency bands. The number of channels in the first and second stages is set to 16 and 32, respectively. In both CSE-Net and FRE-Net, the encoder and decoder contain two repeated Conv blocks with a kernel size of (1, 5) and stride of (1, 2) to reduce the frequency dimension of the features. TAConv is repeated 3 times with a kernel size of (3, 3) and stride (1, 1). The DPRNN is repeated twice in each stage. For the SBF module in the FRE-Net, the kernel size is set to 5. In both the first and second stages, the filter orders of the temporal and frequency deep filters are set to 5.

Our model is trained using the AdamW \cite{adamw} optimizer with a maximum of 100 epochs. The L2 norm for gradient clipping is set to 5, and the learning rate starts at 5e-4, decaying by a factor of 0.98 at the end of each epoch.

\subsection{Ablation study}
To validate the effectiveness of our proposed method, we conducted ablation experiments on the VoiceBank+DEMAND dataset. A set of commonly used metrics was employed for quantitative evaluation, including wideband perceptual evaluation of speech quality (WB-PESQ) \cite{pesq} and three composite metrics \cite{csigcbakcovl} measuring the mean opinion score (MOS): CSIG for signal distortion, CBAK for background noise, and COVL for overall audio quality. We adjusted the number of model channels to maintain comparable model parameters and computational complexity to ensure a fair comparison. 

\begin{table}[!h]
    \caption{Ablation study on VoiceBank+DEMAND test set. M0 represents the model obtained using a single-stage deep filtering method instead of the proposed two-stage hierarchical deep filtering.}
\begin{adjustbox}{width=\columnwidth}
    \renewcommand{\arraystretch}{1.3}
    \label{table:ablation}
    \centering
    \begin{tabular}{@{}l c c c c@{}}
        \toprule
        Model & WB-PESQ$\uparrow$ & CSIG$\uparrow$ & CBAK$\uparrow$ & COVL$\uparrow$ \\
        \midrule
        M0 & 2.97 & 4.22 & 3.46 & 3.60 \\
        HDF-Net (proposed) & \textbf{3.01} & \textbf{4.24} & \textbf{3.52} & \textbf{3.64} \\
        ~~~~w/o SBF & 2.95 & 4.16 & 3.48 & 3.57 \\
        ~~~~w/o TA & 2.92 & 4.12 & 3.41 & 3.53 \\
        
        \bottomrule
    \end{tabular}
\end{adjustbox}
\end{table}

\begin{table}[!h]
\renewcommand{\arraystretch}{1.3}
\caption{Comparison of the model using the proposed temporal deep filtering (TDF) and frequency deep filtering (FDF) methods with the complex ratio mask (CRM).}
\label{table:ablation_df}
\begin{adjustbox}{width=\columnwidth}
\begin{tabular}{@{}ccccccc@{}}
\toprule
Model   & Stage 1 & \multicolumn{1}{l}{Stage 2} & \multicolumn{1}{l}{WB-PESQ$\uparrow$} & CSIG$\uparrow$                     & CBAK$\uparrow$ & COVL$\uparrow$                     \\ \midrule
M1       & CRM    & CRM                         & 2.93                        & 4.15                     & 3.26 & 3.55                     \\
M2       & CRM    & FDF                         & 2.95                        & 4.13                     & 3.43 & 3.54                     \\
M3       & TDF    & CRM                         & 2.97                        & 4.19                        & 3.49    & 3.60                        \\
M4       & TDF    & TDF                         & 2.96                        & 4.18                        & 3.49    & 3.58                        \\ 
M5       & FDF    & TDF                         & 2.99                        & 4.21                     & 3.51 & 3.61                      \\   
HDF-Net (proposed) & TDF    & FDF                 & \textbf{3.01}                        & \textbf{4.24}       & \textbf{3.52} & \textbf{3.64} \\             
\midrule
\end{tabular}
\end{adjustbox}
\end{table}

As shown in Table \ref{table:ablation}, after removing the SBF or TA modules, the performance of the model across all metrics significantly decreases, which demonstrates the effectiveness of subband processing and temporal correlation capture. Additionally, to validate the effectiveness of our proposed hierarchical deep filtering method, we use FRE-Net to output the filter coefficients and apply the deep filtering method from Equation \ref{equation:df} to obtain the results for M0 in Table \ref{table:ablation}. The performance of M0 shows a slight decrease in CSIG compared to the proposed hierarchical deep filtering method, with more marked declines in CBAK and COVL. These results indicate that our hierarchical deep filtering method primarily improves the listening experience by reducing background noise.


As shown in Table \ref{table:ablation_df}, we compare the performance of temporal deep filtering (TDF), frequency deep filtering (FDF), and complex ratio mask (CRM)  methods. In M1, where the CRM method is applied in both stages, we observe a significant performance drop, highlighting the advantage of the proposed method over CRM. M2 and M3 compare the results of using only FDF and TDF. We observe similar results to \cite{deepfiltering,dforderbigworse}, where TDF outperforms FDF. To further exploit TDF, in M4, we apply TDF in both stages, but this leads to a slight performance decrease compared to M3, suggesting that excessive TDF may be unnecessary. In M5, we swap the order of TDF and FDF, resulting in a slight performance drop compared to the proposed model. This indicates that integrating the sub-band processing with frequency deep filtering in the second stage is effective.

\subsection{Comparison with previous advanced systems}

\begin{table}[ht]
    \renewcommand{\arraystretch}{1.3}
    \caption{Performance comparison with previous advanced systems.} 
    \label{table:advancedmodel}
    \begin{adjustbox}{width=\columnwidth}
    \begin{tabular}{@{}ccccccc@{}}
        \toprule
        Model & \begin{tabular}[c]{@{}c@{}}Para.\\ (M)$\downarrow$\end{tabular} & \begin{tabular}[c]{@{}c@{}}MACs\\ (G/s)$\downarrow$\end{tabular} & \begin{tabular}[c]{@{}c@{}}WB-\\ PESQ$\uparrow$\end{tabular} & CSIG$\uparrow$ & CBAK$\uparrow$ & COVL$\uparrow$ \\
        \midrule
        Noisy & - & - & 1.97 & 3.35 & 2.44 & 2.63 \\
        RNNoise \cite{rnnoise} & 0.06 & 0.04 & 2.34 & 3.40 & 2.51 & 2.84 \\
        DeepFilterNet \cite{deepfilternet} & 1.78 & 0.35 & 2.81 & 4.14 & 3.31 & 3.46 \\
        DCCRN \cite{dccrn} & 3.67 & 14.06 & 2.68 & 3.88 & 3.18 & 3.27 \\
        FullSubNet+ \cite{fullsubnet+} & 8.67 & 30.06 & 2.88 & 3.86 & 3.42 &3.57 \\
        CompNet \cite{compnet} & 4.26 & 5.92 & 2.90 & 4.16 & 3.37 & 3.53 \\
        CTS-Net \cite{twoheadsbetterthanone} & 4.35 & 5.57 & 2.92 & 4.25 & 3.46 & 3.59 \\
        DEMUCS \cite{demucs} & 18.87 & 4.32 & 2.93 & 4.22 & 3.25 & 3.52 \\
        PHASEN \cite{phasen} & 8.76 & 6.12 & 2.99 & 4.21 & \textbf{3.55} & 3.62 \\
        GaGNet \cite{Gagnet} & 5.94 & 1.63 & 2.94 & \textbf{4.26} & 3.45 & 3.59 \\
        \midrule
        HDF-Net (proposed) & 0.20 & 0.43 & \textbf{3.01} & 4.24 & 3.52 & \textbf{3.64} \\
        \bottomrule
    \end{tabular}
    \end{adjustbox}
\end{table}

As shown in Table \ref{table:advancedmodel}, we compared the proposed model with other advanced models in terms of parameter size, MACs, and objective metrics on the VoiceBank+DEMAND dataset. 
RNNoise has an extremely small number of parameters and computational complexity, but its performance is modest. Compared to DeepFilterNet, HDF-Net achieve better performance with fewer parameters. 
CTS-Net also adopts a two-stage pipeline and outperforms the proposed model in terms of CSIG. This may be because we use ERB bands to reduce model parameters, resulting in a coarse first-stage output and increasing the burden on the second-stage refinement. 
The proposed HDF-Net achieves comparable performance to other systems, with fewer parameters and lower computational cost.

\section{CONCLUSION}
In this paper, we propose HDF-Net, a two-stage network that decoupled deep filtering, designed to enhance model performance by attending to the information of surrounding TF bins. We also introduce the TAConv module, which further captures temporal correlations for speech enhancement tasks. In addition, we employ several techniques to reduce the model’s computational complexity, enabling the model to achieve competitive performance with a relatively small number of parameters. Our future goal is to validate the applicability of the proposed method to other tasks in the field of speech enhancement.
\newpage
\section{Acknowledgements}
This work was supported in part by the National Natural Science Foundation of China under Grants 62276220 and 62371407, and the Innovation of Policing Science and Technology, Fujian province (Grant number: 2024Y0068)

\bibliographystyle{IEEEtran}
\bibliography{mybib}

\begin{thebibliography}{10}
\providecommand{\url}[1]{#1}
\csname url@samestyle\endcsname
\providecommand{\newblock}{\relax}
\providecommand{\bibinfo}[2]{#2}
\providecommand{\BIBentrySTDinterwordspacing}{\spaceskip=0pt\relax}
\providecommand{\BIBentryALTinterwordstretchfactor}{4}
\providecommand{\BIBentryALTinterwordspacing}{\spaceskip=\fontdimen2\font plus
\BIBentryALTinterwordstretchfactor\fontdimen3\font minus \fontdimen4\font\relax}
\providecommand{\BIBforeignlanguage}[2]{{%
\expandafter\ifx\csname l@#1\endcsname\relax
\typeout{** WARNING: IEEEtran.bst: No hyphenation pattern has been}%
\typeout{** loaded for the language `#1'. Using the pattern for}%
\typeout{** the default language instead.}%
\else
\language=\csname l@#1\endcsname
\fi
#2}}
\providecommand{\BIBdecl}{\relax}
\BIBdecl

\bibitem{deepfilternet}
H.~Schroter, A.~N. Escalante-B, T.~Rosenkranz, and A.~Maier, ``Deepfilternet: A low complexity speech enhancement framework for full-band audio based on deep filtering,'' in \emph{ICASSP 2022 - 2022 IEEE International Conference on Acoustics, Speech and Signal Processing (ICASSP)}, 2022, pp. 7407--7411.

\bibitem{cmgan}
R.~Cao, S.~Abdulatif, and B.~Yang, ``Cmgan: Conformer-based metric gan for speech enhancement,'' in \emph{Interspeech 2022}, 2022, pp. 936--940.

\bibitem{fullsubnet}
X.~Hao, X.~Su, R.~Horaud, and X.~Li, ``Fullsubnet: A full-band and sub-band fusion model for real-time single-channel speech enhancement,'' in \emph{ICASSP 2021 - 2021 IEEE International Conference on Acoustics, Speech and Signal Processing (ICASSP)}, 2021, pp. 6633--6637.

\bibitem{gtcrn}
X.~Rong, T.~Sun, X.~Zhang, Y.~Hu, C.~Zhu, and J.~Lu, ``Gtcrn: A speech enhancement model requiring ultralow computational resources,'' in \emph{ICASSP 2024 - 2024 IEEE International Conference on Acoustics, Speech and Signal Processing (ICASSP)}, 2024, pp. 971--975.

\bibitem{deepfiltering}
W.~Mack and E.~A.~P. Habets, ``Deep filtering: Signal extraction and reconstruction using complex time-frequency filters,'' \emph{IEEE Signal Processing Letters}, vol.~27, pp. 61--65, 2020.

\bibitem{dforderbigworse}
Z.~Zhang, Y.~Xu, M.~Yu, S.-X. Zhang, L.~Chen, D.~S. Williamson, and D.~Yu, ``Multi-channel multi-frame adl-mvdr for target speech separation,'' \emph{IEEE/ACM Transactions on Audio, Speech, and Language Processing}, vol.~29, pp. 3526--3540, 2021.

\bibitem{atwostage}
Y.~Zhang, H.~Zou, and J.~Zhu, ``A two-stage framework in cross-spectrum domain for real-time speech enhancement,'' in \emph{ICASSP 2024 - 2024 IEEE International Conference on Acoustics, Speech and Signal Processing (ICASSP)}, 2024, pp. 12\,587--12\,591.

\bibitem{twoheadsbetterthanone}
A.~Li, W.~Liu, C.~Zheng, C.~Fan, and X.~Li, ``Two heads are better than one: A two-stage complex spectral mapping approach for monaural speech enhancement,'' \emph{IEEE/ACM Transactions on Audio, Speech, and Language Processing}, vol.~29, pp. 1829--1843, 2021.

\bibitem{firstcoarsefineafterward}
F.~Dang, H.~Chen, Q.~Hu, P.~Zhang, and Y.~Yan, ``First coarse, fine afterward: A lightweight two-stage complex approach for monaural speech enhancement,'' \emph{Speech Communication}, vol. 146, pp. 32--44, 2023.

\bibitem{snr1}
Z.~Zheng, Y.~Liu, J.~Liu, K.~Niu, and Z.~He, ``Dual-path transformer based on efficient channel attention mechanism for speech enhancement,'' in \emph{2023 International Conference on Wireless Communications and Signal Processing (WCSP)}, 2023, pp. 7--12.

\bibitem{snr2}
Z.~Zhang, C.~He, S.~Xu, and M.~Wang, ``Real and imaginary part interaction network for monaural speech enhancement and de-reverberation,'' in \emph{2023 Asia Pacific Signal and Information Processing Association Annual Summit and Conference (APSIPA ASC)}, 2023, pp. 972--977.

\bibitem{pconv}
B.-S. Hua, M.-K. Tran, and S.-K. Yeung, ``Pointwise convolutional neural networks,'' in \emph{Proceedings of the IEEE conference on computer vision and pattern recognition}, 2018, pp. 984--993.

\bibitem{rnnoise}
J.-M. Valin, ``A hybrid dsp/deep learning approach to real-time full-band speech enhancement,'' in \emph{2018 IEEE 20th International Workshop on Multimedia Signal Processing (MMSP)}, 2018, pp. 1--5.

\bibitem{erb}
B.~C. Moore, \emph{An introduction to the psychology of hearing}.\hskip 1em plus 0.5em minus 0.4em\relax Brill, 2012.

\bibitem{crn}
K.~Tan and D.~Wang, ``A convolutional recurrent neural network for real-time speech enhancement,'' in \emph{Interspeech 2018}, 2018, pp. 3229--3233.

\bibitem{dprnn}
Y.~Luo, Z.~Chen, and T.~Yoshioka, ``Dual-path rnn: efficient long sequence modeling for time-domain single-channel speech separation,'' in \emph{ICASSP 2020-2020 IEEE International Conference on Acoustics, Speech and Signal Processing (ICASSP)}.\hskip 1em plus 0.5em minus 0.4em\relax IEEE, 2020, pp. 46--50.

\bibitem{dpcrn}
X.~Le, H.~Chen, K.~Chen, and J.~Lu, ``Dpcrn: Dual-path convolution recurrent network for single channel speech enhancement,'' in \emph{Interspeech 2021}, 2021, pp. 2811--2815.

\bibitem{tfattention}
Q.~Zhang, Q.~Song, Z.~Ni, A.~Nicolson, and H.~Li, ``Time-frequency attention for monaural speech enhancement,'' in \emph{ICASSP 2022 - 2022 IEEE International Conference on Acoustics, Speech and Signal Processing (ICASSP)}, 2022, pp. 7852--7856.

\bibitem{sequeezeandexcitation}
J.~Hu, L.~Shen, and G.~Sun, ``Squeeze-and-excitation networks,'' in \emph{Proceedings of the IEEE conference on computer vision and pattern recognition}, 2018, pp. 7132--7141.

\bibitem{voicebank_demand}
C.~V. Botinhao, X.~Wang, S.~Takaki, and J.~Yamagishi, ``Investigating rnn-based speech enhancement methods for noise-robust text-to-speech,'' in \emph{9th ISCA Speech Synthesis Workshop}, 2016, pp. 159--165.

\bibitem{adamw}
I.~Loshchilov, ``Decoupled weight decay regularization,'' \emph{arXiv preprint arXiv:1711.05101}, 2017.

\bibitem{pesq}
A.~W. Rix, J.~G. Beerends, M.~P. Hollier, and A.~P. Hekstra, ``Perceptual evaluation of speech quality (pesq)-a new method for speech quality assessment of telephone networks and codecs,'' in \emph{2001 IEEE international conference on acoustics, speech, and signal processing. Proceedings (Cat. No. 01CH37221)}, vol.~2.\hskip 1em plus 0.5em minus 0.4em\relax IEEE, 2001, pp. 749--752.

\bibitem{csigcbakcovl}
Y.~Hu and P.~C. Loizou, ``Evaluation of objective quality measures for speech enhancement,'' \emph{IEEE Transactions on audio, speech, and language processing}, vol.~16, no.~1, pp. 229--238, 2007.

\bibitem{dccrn}
Y.~Hu, Y.~Liu, S.~Lv, M.~Xing, S.~Zhang, Y.~Fu, J.~Wu, B.~Zhang, and L.~Xie, ``Dccrn: Deep complex convolution recurrent network for phase-aware speech enhancement,'' \emph{Interspeech 2020}, 2020.

\bibitem{fullsubnet+}
J.~Chen, Z.~Wang, D.~Tuo, Z.~Wu, S.~Kang, and H.~Meng, ``Fullsubnet+: Channel attention fullsubnet with complex spectrograms for speech enhancement,'' in \emph{ICASSP 2022-2022 IEEE International Conference on Acoustics, Speech and Signal Processing (ICASSP)}.\hskip 1em plus 0.5em minus 0.4em\relax IEEE, 2022, pp. 7857--7861.

\bibitem{compnet}
X.~Liang, J.~Yang, G.~Lu, and D.~Zhang, ``Compnet: Competitive neural network for palmprint recognition using learnable gabor kernels,'' \emph{IEEE Signal Processing Letters}, vol.~28, pp. 1739--1743, 2021.

\bibitem{demucs}
A.~Défossez, G.~Synnaeve, and Y.~Adi, ``Real time speech enhancement in the waveform domain,'' in \emph{Interspeech 2020}, 2020, pp. 3291--3295.

\bibitem{phasen}
D.~Yin, C.~Luo, Z.~Xiong, and W.~Zeng, ``Phasen: A phase-and-harmonics-aware speech enhancement network,'' in \emph{Proceedings of the AAAI Conference on Artificial Intelligence}, vol.~34, no.~05, 2020, pp. 9458--9465.

\bibitem{Gagnet}
A.~Li, C.~Zheng, L.~Zhang, and X.~Li, ``Glance and gaze: A collaborative learning framework for single-channel speech enhancement,'' \emph{Applied Acoustics}, vol. 187, p. 108499, 2022.

\end{thebibliography}

\end{document}